\documentclass[aps,pre,10pt,showpacs,twocolumn,nofootinbib,floatfix]{revtex4}
\usepackage{graphicx,amsmath,amssymb}
\usepackage[usenames]{color}
\usepackage[latin1]{inputenc}
\usepackage{ulem}

\newcommand{\eq}[1]{Eq.~(\ref{#1})}
\newcommand{\fig}[1]{Fig.~\ref{#1}}

\newcommand{\sect}[1]{Section~\ref{#1}}
\newcommand{\avg}[1]{\langle #1 \rangle}
\newcommand{\olcite}[1]{Ref.~\onlinecite{#1}}
\newcommand{\ahum}[1]{``#1''}
\newcommand{\comment}[1]{ }
\newcommand{\LV}{liquid-vapor }

\begin{document}


\title{Phase separation in fluids exposed to spatially periodic external fields}

\author{R. L. C. Vink}
\affiliation{Institute of Theoretical Physics, Georg-August-Universit\"at 
G\"ottingen, Friedrich-Hund-Platz~1, D-37077 G\"ottingen, Germany}

\author{A. J. Archer}
\affiliation{Department of Mathematical Sciences, Loughborough University, 
Loughborough, Leicestershire, LE11 3TU, United Kingdom}

\begin{abstract} We consider the \LV type phase transition for fluids confined within 
spatially periodic external fields. For a fluid in $d=3$ dimensions, the 
periodic field induces an additional phase, characterized by large density 
modulations along the field direction. At the triple point, all three phases 
(modulated, vapor, and liquid) coexist. At temperatures slightly above the 
triple point and for low (high) values of the chemical potential, two-phase coexistence 
between the modulated phase and the vapor (liquid) is observed. We study this 
phenomenon using computer simulations and mean-field theory for the Ising model. 
The theory shows that, in order for the modulated phase to arise, the field 
wavelength must exceed a threshold value. We also find an extremely low tension 
of the interface between the modulated phase and the vapor/liquid phases. The 
tension is of the order $10^{-4} \, k_BT$ per squared lattice spacing, where 
$k_B$ is the Boltzmann constant, and $T$ the temperature. In order to detect such 
low tensions, a new simulation method is proposed. We also consider the case of 
$d=2$ dimensions. The modulated phase then does not survive, leading to a 
radically different phase diagram. \end{abstract}

\maketitle

\section{Introduction}

Liquid-vapor type phase transitions in fluids are profoundly affected by confinement 
(for a recent review see \olcite{citeulike:3533972}). Typical effects are the 
depression of critical temperatures \cite{fisher.nakanishi:1981}, changes in 
universality \cite{gennes:1984}, or entirely new phenomena altogether 
\cite{citeulike:7749811, citeulike:9633515}. The confinement of a fluid between 
two parallel surfaces is arguably the most simple example one could envision 
\cite{fisher.nakanishi:1981}. Already for this case the corresponding phase 
behavior is extremely rich, especially if the surfaces have different 
interactions with the fluid \cite{citeulike:9633515, virgiliis.vink.ea:2006}. 
With the advance of microcontact printing \cite{citeulike:9608754}, vapor 
deposition and grafting methods \cite{citeulike:9726164}, as well as 
photolithography \cite{citeulike:4014772}, the possibilities of tuning the 
surface-fluid interaction are essentially endless. In addition to surfaces, 
confinement in fluids may also be induced via external fields (for example, 
optical tweezers can be used to realize one-dimensional diffusion channels for 
colloidal particles in suspension \cite{citeulike:9609811}). Hence, 
well-characterized geometries of ever increasing complexity can be generated, 
and the phase behavior of fluids confined within these is expected to become 
correspondingly richer.

With these developments in mind, this paper considers the fate of the \LV 
transition in a fluid confined within a static external field having periodic spatial 
oscillations in one direction. In $d=2$ dimensions, such a field might be 
realized using a stripe-patterned surface \cite{citeulike:4729899, 
citeulike:9258031}, while in $d=3$ dimensions, laser \cite{citeulike:9726153} or 
electric fields \cite{citeulike:5887157, citeulike:9726103, citeulike:9726113} 
could possibly be used. The case $d=3$ was first considered theoretically in 
\olcite{citeulike:7530595} for a colloid-polymer mixture. The main finding was a 
new kind of phase transition, referred to as {\it laser-induced condensation} 
(LIC), which takes place provided the field wavelength is large enough. In the 
presence of the periodic field, one then observes a new third phase (in addition 
to the vapor and liquid phases) characterized by (i) an average density between 
that of the vapor and liquid phase, and (ii) featuring large density modulations 
along the field direction (because of the latter modulations we refer to this 
phase as the \ahum{zebra} phase in what follows). The presence of the zebra 
phase dramatically alters the \LV phase diagram: the critical point of the bulk 
transition is replaced by two new critical points and a triple point. At 
temperatures between the triple and critical points, vapor-zebra and 
liquid-zebra two-phase coexistence is observed (at low and high values of the 
chemical potential, respectively).

In a subsequent publication \cite{citeulike:9633508} the nature of the critical 
points was elucidated, and also the tensions $\gamma_{\rm vz}$ and $\gamma_{\rm 
lz}$ of, respectively, the vapor-zebra and liquid-zebra interfaces were 
calculated. The main observations were a critical behavior corresponding to 
effectively $d=2$ dimensions (i.e.~one below the system dimension), and 
extremely low interfacial tensions. The latter were found, using density functional 
theory, to be at most $\gamma_{\rm vz} \sim \gamma_{\rm lz} \sim 10^{-5}~k_BT$ 
per projected particle area (with $k_B$ the Boltzmann constant, and $T$ the 
temperature). The accompanying simulations confirm that $\gamma_{\rm vz}$ and 
$\gamma_{\rm lz}$ must be extremely low, but no numerical values could be 
obtained (from the simulation data of \olcite{citeulike:9633508}, interface 
tensions of exactly zero cannot be completely ruled out either).

In this paper, we revisit LIC using computer simulations and mean-field theory 
for the Ising model. Compared to a colloid-polymer mixture, computer simulations of the Ising model 
allows for much faster equilibration, such that larger system sizes can be 
reached. In addition, the underlying spin reversal symmetry of the Ising model 
makes the finite-size scaling analysis much more straight-forward. Of course, since the universality 
class of fluids is the Ising one, generic trends observed in the latter directly 
apply to fluids as well. We first consider LIC in $d=3$ dimensions. The 
corresponding phase diagram is calculated using both simulation and theory. In 
particular, we demonstrate how the phase diagram depends on the field wavelength 
and amplitude. Our next aim is to measure the interface tensions $\gamma_{\rm 
vz}$ and $\gamma_{\rm lz}$: it is important to confirm the density functional 
prediction that $\gamma_{\rm vz}$ and $\gamma_{\rm lz}$ are extremely low but 
finite. As it turns out, accurate measurements of such extremely low tensions 
are beyond the scope of \ahum{standard} methods \cite{citeulike:9039746, 
citeulike:9246548}, and so an alternative route is proposed. Finally, we 
consider LIC in $d=2$ dimensions. Since the critical behavior was shown to be of 
dimension $d-1$ \cite{citeulike:9633508}, we expect radical departures from the 
previously considered case $d=3$. Indeed, in $d=2$ dimensions, the two critical 
points do not survive, and an altogether different phase diagram is obtained.

\section{Model and Simulation method}

We consider the Ising model on rectangular $L \times L \times D$ ($d=3$) and $L 
\times D$ ($d=2$) lattices with periodic boundary conditions in all 
directions. The system is exposed to a periodic external field $V_{\rm per}(z)$, 
with the $z$-axes parallel to edge $D$ of the lattice. To each lattice 
site $i$, a spin variable $s_i = \pm 1$ is attached. The energy of the system is 
given by
\begin{equation}
 E = -J \sum_{\avg{i,j}} s_i s_j + H \sum_i s_i + 
 \sum_i s_i V_{\rm per}(z_i),
 \label{eq:Hamiltonian}
\end{equation}
where the first sum is over nearest neighbors, and the remaining sums over 
sites. The first term is the usual Ising pair interaction with coupling 
constant~$J$ (we consider ferromagnetic interactions $J>0$ only). The second 
term is the interaction of the spins with a homogeneous external magnetic field 
of strength~$H$. The last term represents the interaction with the periodic 
field, where $z_i$ is the $z$-coordinate of spin~$i$. For the periodic external field we 
use a block wave of alternating sign
\begin{equation}
V_{\rm per}(z) = \begin{cases}
 -h & 0 < z \leq \lambda/2, \\
 +h & \lambda/2 < z \leq \lambda, \\
\end{cases}
\label{eq:ext_pot}
\end{equation}
with $h$ the field strength, and $\lambda$ the wavelength. Due to the 
discretization of the lattice we must choose $\lambda=2an_1$, with $a$ the 
lattice constant, and $n_1$ an integer. The use of periodic boundary conditions 
implies that the lattice edge $D=\lambda n_2$, with $n_2$ also an integer. In 
what follows, the lattice constant is the unit of length $a \equiv 1$. In 
addition, a factor of $1/k_BT$ is assumed to have been absorbed into the 
coupling constants $J$, $H$ and $h$ such that these quantities are dimensionless.

Monte Carlo simulations and mean-field theory are used to study the phase behavior of the above model. 
The key output of the simulations is the distribution $P(m)$, defined as the 
probability of observing the system in a state with magnetization~$m = (1/N) 
\sum_i s_i$, with $N=DL^{d-1}$ the total number of lattice sites. We emphasize 
that $P(m)$ depends on all the model parameters introduced above, including the 
system size. To obtain $P(m)$ we use single spin-flip dynamics 
\cite{newman.barkema:1999} combined with successive umbrella sampling 
\cite{virnau.muller:2004}; the latter scheme ensures that $P(m)$ is obtained 
over the entire range $-1 \leq m \leq 1$, including those regions where $P(m)$ 
is very small. We also use histogram reweighting \cite{ferrenberg.swendsen:1988} 
to extrapolate data obtained for one set of values of the coupling constants to 
different (nearby) values.

\section{Results in $d=3$ dimensions}

In this section we present results for the case $d=3$. We begin in Sec.\ \ref{subsec:LIC_pd}
with simulation results obtained for an external potential with strength $h=0.075$
and wavelength $\lambda=10$. Following this, in Sec.\ \ref{subsec:MF}, we present
mean-field theory results for how the phase diagram varies as the parameters
$h$ and $\lambda$ are varied.

\subsection{Laser-induced condensation: phase diagram}
\label{subsec:LIC_pd}

\begin{figure*}
\begin{center}
\includegraphics[width=1.5\columnwidth]{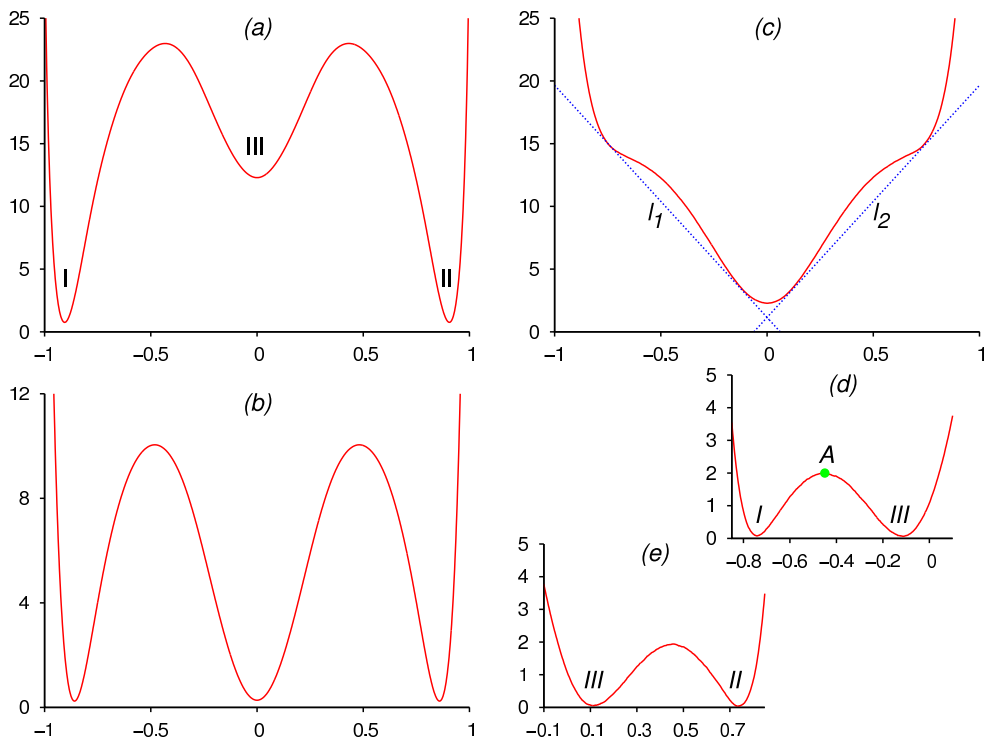}

\caption{\label{fig1} LIC in the $d=3$ Ising model; plotted in each of the 
graphs is the free energy $F(m)$ in units of $k_BT$ (vertical axes) versus the 
magnetization $m$ (horizontal axes). The free energy curves present actual 
simulation data obtained for system size $L=10$ and $D=\lambda$. (a) Free energy 
for $J>J_{\rm tr}$ and $H=0$, i.e.~above the triple point. A coexistence between 
two phases, I and II, is observed. (b) Free energy for $J=J_{\rm tr}$ and $H=0$, 
i.e.~exactly at the triple point; three-phase coexistence is observed. (c) Free 
energy measured between the triple and critical points, $J_{\rm cr} < J < J_{\rm 
tr}$, but still using $H=0$. We now observe two \ahum{common tangent} lines: 
$l_1$ and $l_2$. By choosing $H=\Delta_1$, where $\Delta_1$ is determined from 
the slope of $l_1$, coexistence between phases I and III can be induced (d). 
Similarly, from the slope of $l_2$, we obtain $H=\Delta_2$, at which phases II 
and III coexist (e).}

\end{center}
\end{figure*}

\begin{figure}
\begin{center}
\includegraphics[width=0.6\columnwidth]{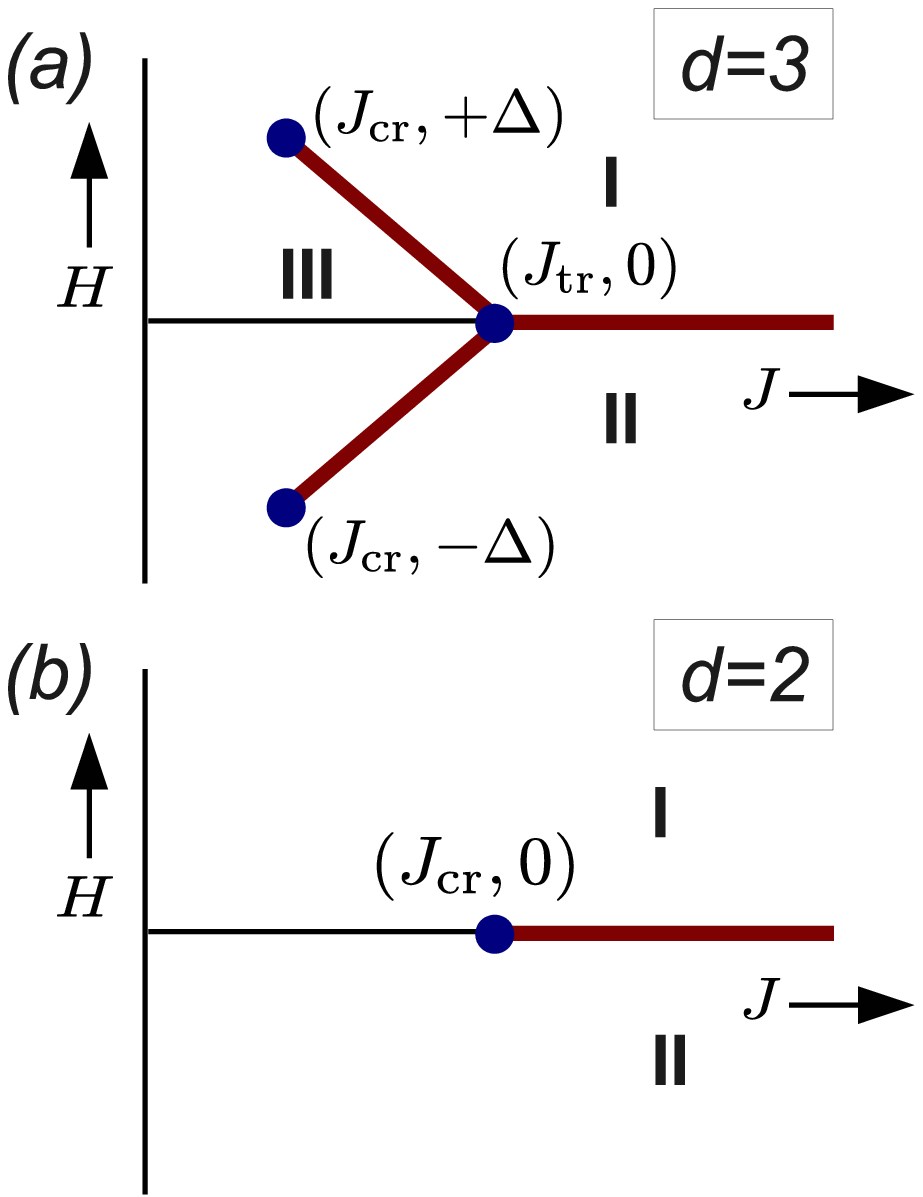}

\caption{\label{fig2} (a) Phase diagram of LIC for the $d=3$ Ising model. The 
phase diagram is a symmetric \ahum{pitchfork} featuring one triple point, and 
two critical points (indicated by dots). The lines correspond to first-order 
phase transitions (between phases I-II, I-III, and II-III). (b) LIC in $d=2$ 
dimensions. In this case, there is no phase III. The phase diagram features only 
a single line of first-order phase transitions terminating in a critical point.}

\end{center}
\end{figure}

To understand LIC in the Ising model it is best to consider the free energy 
$F(m)$ as function of the magnetization~$m$. The latter is related to the 
magnetization probability distribution, $F(m) = - k_B T \ln P(m)$, to which we 
have direct access in our simulations. In \fig{fig1}(a), we show $F(m)$ for a 
high value of the coupling constant~$J$ and $H=0$. The salient features are two 
global minima, at low and high values of $m$, reflecting a coexistence between 
two phases (I and II). We also observe a local minimum at $m=0$, corresponding 
to a phase~III, but it is meta-stable. In (b), we plot $F(m)$ for a lower value 
of $J$ and $H=0$. We now observe three minima at equal height corresponding to a 
triple point, where all three phases coexist. Next, in (c), we show $F(m)$ for 
an even lower value of $J$, but still using $H=0$. There is now only one global 
minimum at $m=0$. However, by applying an appropriate homogeneous field 
$H=\Delta_1$, a coexistence between phase I and III is obtained (d). The value 
of $\Delta_1$ follows from the slope of the \ahum{common tangent} line $l_1$. 
Similarly, by applying a homogeneous field $H=\Delta_2$ (determined from the 
slope of line $l_2$), coexistence between phase II and III is obtained (e). 
Finally, at some critical value $J=J_{\rm cr,1} (J_{\rm cr,2})$, the I-III 
(II-III) coexistence line terminates, below which there is only one phase.

\fig{fig1} is the analogue of LIC \cite{citeulike:7530595} in the Ising model, 
with phase I playing the role of vapor~(v), phase II of the liquid~(l), and 
phase III of the zebra (z) phase. Due to spin reversal symmetry, it holds that 
$H=0$ at the triple point $J=J_{\rm tr}$. Below the triple point, symmetry 
implies that $\Delta_1 = -\Delta_2 \equiv \Delta$ and $J_{\rm cr,1} = J_{\rm 
cr,2} \equiv J_{\rm cr}$. The resulting phase diagram is a symmetric pitchfork 
(\fig{fig2}(a)). The crucial difference with fluids (which typically lack spin 
reversal symmetry) is that the phase diagram is asymmetric in that case: $J_{\rm 
cr,1} \neq J_{\rm cr,2}$ and the fields (chemical potentials) $\Delta_i$ are not 
trivially related to each other \cite{citeulike:9633508}.

We emphasize that the free energy curves in \fig{fig1} are obtained in 
simulations using $D=\lambda$. If one instead uses $D=\lambda n_2$ with integer 
$n_2>1$, one finds that $F(m)$ develops additional minima, as discussed in 
detail in \olcite{citeulike:9633508}. These additional minima reflect 
meta-stable coexistence states and should not be confused with new phases. 
Hence, also when $D>\lambda$, the generic mechanism of LIC as shown in 
\fig{fig1} still applies.

\subsection{Stability of the zebra phase: \\ mean-field calculations}
\label{subsec:MF}

\begin{figure}
\centering
\includegraphics[width=0.85\columnwidth]{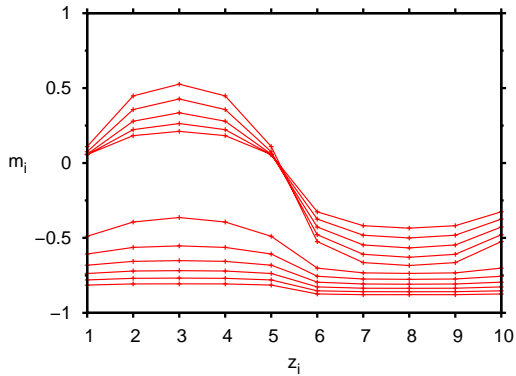} 
\caption{\label{fig:profiles} A sequence of magnetization profiles for 
varying $J$, going from $J=0.14$ (top) to $J=0.24$ (bottom), in increments of 
0.01 obtained from the mean-field theory. For low values of $J$, one observes 
phase III, which is characterized by large modulations in the magnetization 
profile. For high values of $J$, phase~II is observed, which is characterized by 
an overall low value of the magnetization and only small modulations. Note the 
discontinuous change in the profiles as the system crosses the II--III phase 
boundary (here at $J \approx 0.18-0.19$). The remaining parameters used in 
the figure are $H=0.03$, $h=0.075$ and $\lambda=10$.}
\end{figure}

\begin{figure}
\begin{center}
\includegraphics[width=0.85\columnwidth]{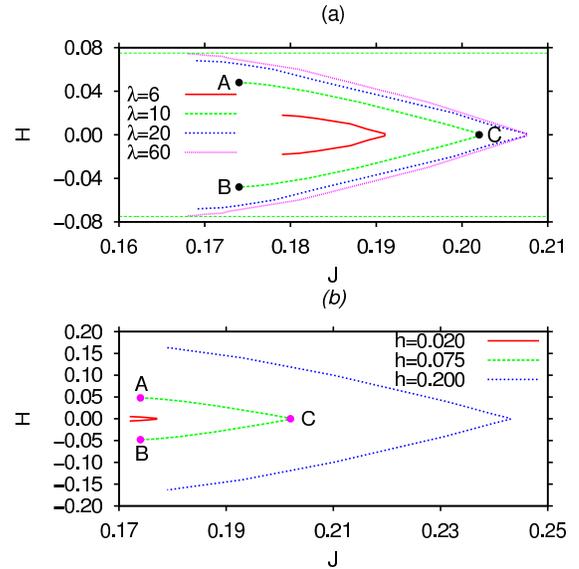}
\caption{\label{fig:bw} LIC phase diagrams for $d=3$ obtained using mean-field 
theory (for clarity only the transition lines between the zebra III phase and 
the I and II phases are shown). In (a) we show results for $h=0.075$ and various 
values of the field wavelength $\lambda$. In (b) we show results for 
$\lambda=10$ and various values of the field amplitude $h$.}
\end{center}
\end{figure}

In order to develop a qualitative understanding of how the LIC phase diagram of 
\fig{fig2}(a) depends on the external field wavelength $\lambda$ and amplitude 
$h$, we use the following simple mean-field (Bragg-Williams) approximation 
\cite{PlBi06,ChLu00} for the free energy $F$ of the system with Hamiltonian $E$, 
given in Eq.\ \eqref{eq:Hamiltonian}:
\begin{eqnarray}\notag
F=\sum_i\Bigg[ k_BT\frac{1+m_i}{2}\ln\left(\frac{1+m_i}{2}\right)\\ \notag
+k_BT\frac{1-m_i}{2}\ln\left(\frac{1-m_i}{2}\right) \\
+Hm_i-m_iV_{\rm per}(z_i)\Bigg]-J\sum_{\langle i,j\rangle}m_im_j,
\label{eq:F_BW}
\end{eqnarray}
where $m_i\equiv\langle s_i\rangle$ is the average magnetization at lattice site 
$i$. For a given external potential $V_{\rm per}(z)$, the average magnetization 
profile corresponds to the set $\{m_1,m_2,\cdots\}$ which minimize the free 
energy \eqref{eq:F_BW}; i.e.\ are the solution to the set of equations $\partial 
F/\partial m_i=0$. This yields the following set of $i$ simultaneous equations:
\begin{eqnarray}
\frac{k_BT}{2}\ln\left(\frac{1+m_i}{1-m_i}\right)+H-V_{\rm per}(z_i)-J\sum'_j m_j=0,
\label{eq:EL}
\end{eqnarray}
where $\sum'_j$ denotes the sum over the 6 (in $d=3$) nearest neighbor lattice 
sites of site $i$. Because the external potential $V_{\rm per}(z)$ in Eq.\ 
\eqref{eq:ext_pot}, only varies in the $z$-direction, we have magnetization 
profiles that only vary in this one direction and so solving Eqs.\ \eqref{eq:EL} 
is straightforward. We do so using a simple (Picard) iterative numerical scheme. 
In Fig.\ \ref{fig:profiles} we show some example magnetization profiles 
calculated for various different values of $J$ as one crosses the transition 
line from phase II to the zebra phase III. We see a discontinuous change in the 
average magnetization in the system as one crosses the phase transition.

In \fig{fig:bw}(a) we show phase diagrams for various values of the field 
wavelength $\lambda$ and fixed field amplitude $h=0.075$ (the upper and lower 
horizontal lines correspond to $\pm h$, respectively). Note that for clarity we 
only display the I-III and II-III coexistence lines and do not display the I-II 
\LV coexistence line. For $\lambda=10$, the zebra critical points are marked $A$ 
and $B$, while point $C$ indicates the triple point. In the limit $\lambda \to 
\infty$, the critical points $A$ and $B$ shift toward $(J_{\rm cr,bulk},\pm h)$, 
respectively, where $J_{\rm cr,bulk}=1/6$ in the mean-field theory. As $\lambda 
\to \infty$, essentially two infinite systems are obtained: one inside a 
positive (homogeneous) external field $h$, and one inside a negative field $-h$. 
The value of $H$ at the respective critical point simply has to \ahum{cancel} 
this field. In the opposite limit $\lambda \to \lambda_{\rm min}=2$, we observe 
the loss of the zebra phase. In order for the zebra phase to survive, 
$\lambda/2$ must exceed the bulk correlation length which is the quantity that 
determines the distance over which the average density changes from one value to 
another. Indeed, for $\lambda =4$ and smaller, the critical points $A$ and $B$ 
can no longer be identified, and only point $C$ survives (which then no longer 
is a triple point, but a critical point, marking the end of the I-II coexistence 
region). When $\lambda=4$ and $h=0.075$ the mean-field critical point is at 
$J\approx0.177$ and when $\lambda=2$ it is at $J\approx0.169$. Recall that the 
mean-field bulk critical point (i.e.\ for $h=0$) is at $J_{\rm 
cr,bulk}=1/6\approx0.167$.

In \fig{fig:bw}(b) we show phase diagrams for fixed $\lambda=10$ (chosen above 
the threshold such that the zebra phase survives) and various values of the 
field amplitude $h$. In the limit $h \to 0$, we observe that the points $A$, $B$ 
and $C$ all approach $(J_{\rm cr,bulk}=1/6,H=0)$, the critical point of the bulk 
system. When $h$ is very small, it is difficult to locate numerically the 
transition points. However, a threshold value of $h$ below which the zebra phase 
vanishes appears to be absent in this case (in contrast to the case as $\lambda$ 
is decreased). The effect of increasing $h$ is that the I-III and II-III 
transition lines open-up, with the transition points $A$, $B$ and $C$ shifting 
toward larger values of $J$. Note that the value of $H$ at the critical point is 
always less in magnitude than the value of $h$. When $h=0.02$, then the critical 
value $H_{\rm cr}\approx 0.005$; when $h=0.2$, then $H_{\rm cr}\approx 0.163$ 
and when $h=2$, then $H_{\rm cr}\approx 1.957$. We see from these that as $h$ 
becomes large, then $H_{\rm cr} \to h$.

\subsection{Finite-size scaling analysis}
\label{sec:fss}

We now continue with our simulation analysis using $\lambda=10$ and $h=0.075$. 
Finite-size scaling is used to locate the triple and critical points. We measure 
$P(m)$ for various values of~$L$, keeping $D=\lambda$ fixed. We thus assume that 
correlations in the $z$-direction are \ahum{cut-off} by the periodic field, and 
so we do not need to scale in this direction (we return to this point shortly). 
The distribution $P(m)$ is always measured at $H=0$ and symmetrized by hand 
afterward such that $P(m)=P(-m)$, thereby imposing the spin reversal symmetry of 
the Ising model; subsequent histogram reweighting (in $J$ and $H$) is performed 
using the symmetrized distribution.

\begin{figure}
\begin{center}
\includegraphics[width=0.9\columnwidth]{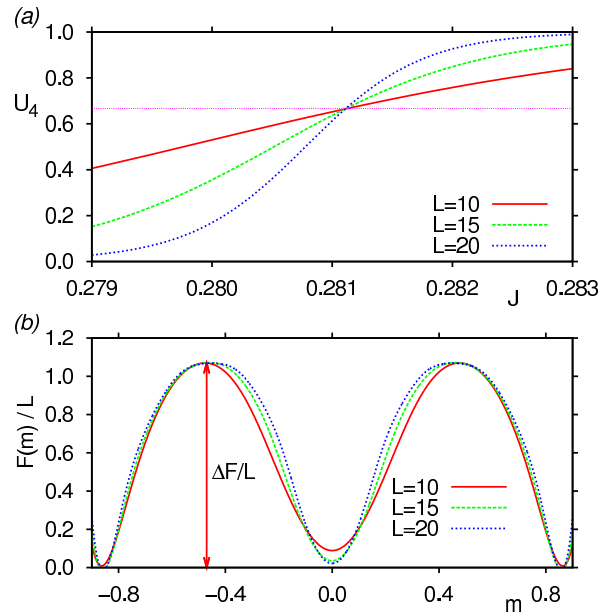}

\caption{\label{fss_tr} Finite-size scaling analysis to locate the LIC triple 
point for the $d=3$ Ising model. (a) Binder cumulant $U_4$ versus the coupling 
constant $J$ for various system sizes $L$. The curves strikingly intersect at 
$U_4=2/3$ (horizontal line) of a triple point; the value of $J$ at the 
intersection yields $J_{\rm tr}$. (b) The {\it scaled} free energy $F(m)/L$ 
precisely at the triple point, $J=J_{\rm tr}$, for various system sizes. Clearly 
visible are the three minima, corresponding to the coexisting phases, and a free 
energy barrier that increases linearly with $L$.}

\end{center}
\end{figure}

To determine $J_{\rm tr}$ we use $H=0$ and assume that, precisely at the triple 
point, the magnetization distribution $P(m)$ is a superposition of three 
(non-overlapping) Gaussian peaks, centered around $m=-m_0$, $m=0$, and $m=+m_0$, 
respectively. For such a triple-peaked distribution one may easily show that the 
Binder cumulant $U_4=2/3$, with the cumulant defined as
\begin{equation}
 U_4 \equiv \avg{m^2}^2 / \avg{m^4}, \quad 
 \avg{m^k} = \int_{-1}^{+1} m^k P(m) \, dm,
\end{equation}
and where it is assumed that $P(m)$ is normalized. In \fig{fss_tr}(a), for the 
case when $h=0.075$ and $\lambda=10$ we plot $U_4$ versus $J$ for various system 
sizes $L$. The curves strikingly intersect at the expected \ahum{height} of a 
triple point, from which we conclude that $J_{\rm tr} \approx 0.2811$. Note that 
this exceeds $J_{\rm cr,bulk} \approx 0.2217$ of the critical point in the bulk 
($h=0$) Ising model \cite{orkoulas.panagiotopoulos.ea:2000}, consistent with the 
general observation that confinement lowers transition temperatures. In 
\fig{fss_tr}(b), we plot the free energy $F(m)$ at the triple point for various 
system sizes. The curves clearly show the three minima of the coexisting phases. 
Note that the minima are shifted to zero, and that the vertical scale is divided 
by~$L$. In this representation, the barrier $\Delta F/L$ (vertical arrow) is 
approximately constant. Hence, at the triple point, we observe a free energy 
barrier that increases linearly with the system size $\Delta F \propto L$. This 
implies that the general shape of $F(m)$, i.e.~featuring three minima, persists 
in the thermodynamic limit $L \to \infty$, and thus reflects a genuine triple 
point (see also \olcite{citeulike:3908342} where these ideas were first applied 
to first-order phase transitions).

\begin{figure}
\begin{center}
\includegraphics[width=0.9\columnwidth]{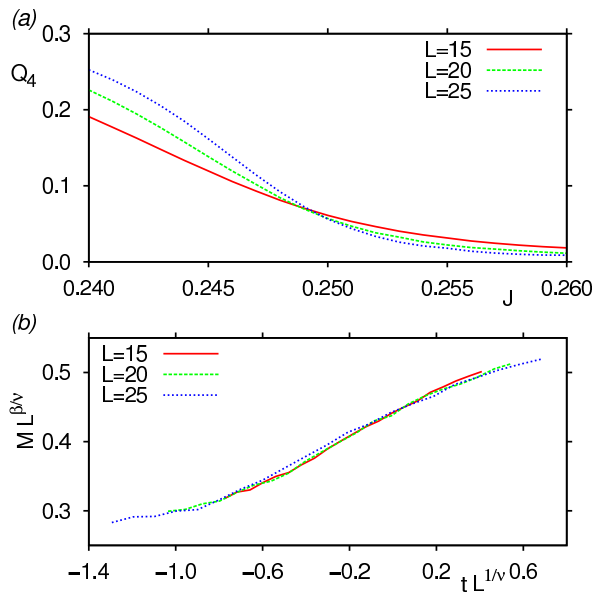}
\caption{\label{fss_cr} Finite-size scaling analysis to locate the LIC critical 
point for the $d=3$ Ising model. (a) Cumulant $Q_4$ versus the coupling constant 
$J$ for various system sizes $L$; the intersection yields $J=J_{\rm cr}$. (b) 
Scaling plot of the order parameter $M$ where $t=J/J_{\rm cr}-1$; by using 
$J_{\rm cr} \approx 0.2531$ and $d=2$ Ising values for the critical exponents 
$\beta,\nu$ the data for different $L$ collapse onto a single master curve.}
\end{center}
\end{figure}

For values of $J$ between the triple and critical points, coexistence with the 
zebra phase (phase~III) is observed at appropriate values $H=\pm \Delta$ of the 
external magnetic field. To locate $J_{\rm cr}$ we perform the same cumulant 
analysis as in our previous work \cite{citeulike:9633508}. For a given value of 
$J$, $\avg{m}$ and $U_4$ are measured as function of $H$ (due to symmetry, only 
$H \geq 0$ needs to be considered). One then uses these data to construct a 
graph of $U_4$ versus $\avg{m}$ (which thus is parametrized by $H$). The 
resulting curve reveals a maximum, corresponding to $H=\Delta$, enveloped by two 
minima \cite{kim.fisher:2004}. The average value of the cumulant at the minima 
equals~$Q_4$. In \fig{fss_cr}(a), we plot $Q_4$ versus $J$ for different $L$; 
from the intersection point we conclude that $J_{\rm cr} \approx 0.2531$ (the 
corresponding critical field $\Delta_{\rm cr} \approx 0.017$). The difference 
$M$ in the magnetizations $\avg{m}$ at the minima yields the order parameter. 
The latter is analyzed in the finite-size scaling plot 
\cite{newman.barkema:1999, binder.heermann:2002} of \fig{fss_cr}(b). The key 
result is that, by using the $d=2$ Ising values for the critical exponents 
($\beta=1/8, \nu=1$) the data for different $L$ collapse. The critical points of 
LIC in $d=3$ dimensions thus remain in the universality class of the Ising 
model, but of reduced dimensionality $d-1$. We believe that colloid-polymer 
mixtures should ultimately yield the same exponents, but that their complexity 
still prevents efficient simulations of large enough systems to explicitly see 
this \cite{citeulike:9633508}.

\subsection{Measurement of ultra-low surface tension}

We now consider the regime between the triple and critical points using $H=\pm 
\Delta$. The free energy then schematically resembles \fig{fig1}(d,e), 
corresponding to I-III and II-III phase coexistence, respectively. Hence, there 
will be interfaces present, and our aim is to measure the corresponding 
interface tension $\gamma$ (due to symmetry it holds that 
$\gamma_{I,III}=\gamma_{II,III} \equiv \gamma$, of course). Following density 
functional calculations \cite{citeulike:9633508}, $\gamma$ is expected to be 
extremely low. In principle, for \LV transitions, the corresponding interface 
tension $\gamma_{\rm lv}$ can be accurately determined from the free energy 
using an idea of Binder \cite{binder:1982}. In our previous work 
\cite{citeulike:9633508}, we discussed how this approach may be generalized to 
LIC, but it was clear that present computer power is not sufficient to reach the 
system sizes required for this method to work. Hence, we propose a different 
method.

\begin{figure}
\begin{center}
\includegraphics[width=0.9\columnwidth]{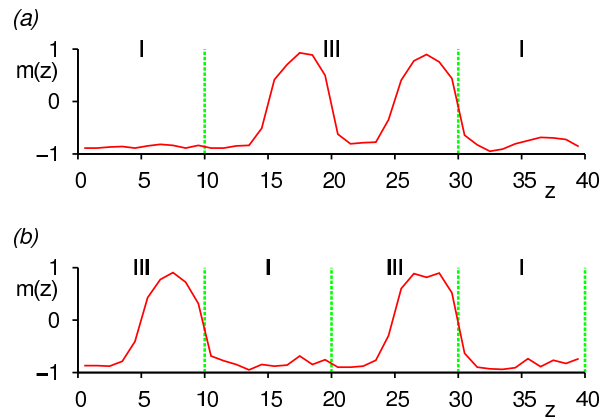}
\caption{\label{prof} Instantaneous magnetization profiles $m(z)$ obtained for 
$J=0.275$, $L=14$, and fixed overall magnetization $m=-0.426$. In (a) we observe 
a coexistence between two domains, while in (b) a coexistence between four 
domains is seen. The vertical lines indicate the approximate locations of the 
interfaces. By counting how often the arrangements (a) and (b) occur during a 
long simulation run, the interface tension can be determined, see \eq{fit}.}
\end{center}
\end{figure}

\begin{figure}
\begin{center}
\includegraphics[width=0.9\columnwidth]{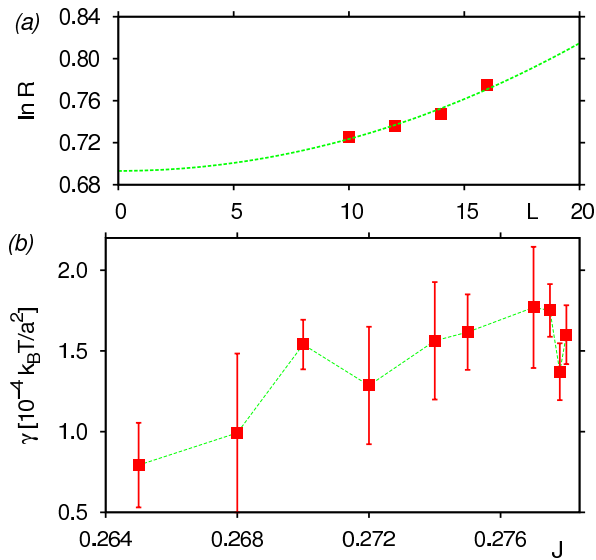}

\caption{\label{gam} (a) Variation of $\ln R$ with system size $L$ for 
$J=0.275$; symbols are simulation data, the curve is a fit to \eq{fit}. Note 
that the data are consistent with $\lim_{L \to 0} \ln R = \ln 2$ of the entropy 
difference. (b) The interfacial tension $\gamma$, in units of $k_BT$ per squared 
lattice spacing, as a function of $J$. In agreement with theoretical expectations 
\cite{citeulike:9633508}, $\gamma$ is extremely low and decreases as $J \to 
J_{\rm cr}$.}

\end{center}
\end{figure}

The key observation is that the periodic field $V_{\rm per}(z)$ suppresses 
interface fluctuations (capillary waves) in the $z$-direction: even though 
$\gamma$ is very low, the I-III and II-III interfaces are sharp. This is in 
contrast to conventional \LV interfaces which, at low interface tension, are 
extremely broad \cite{vink.horbach.ea:2004*b, aarts.schmidt.ea:2004}. The fact 
that the interfaces remain sharp is the property we exploit to extract~$\gamma$. 
To this end, we consider a simulation box with edge $D=4\lambda$. In \fig{prof}, 
we show instantaneous magnetization profiles $m(z)$ obtained for two 
equilibrated samples at {\it fixed} overall magnetization~$m$. The value of $m$ 
must be chosen such that half the system is occupied by phase~I, and the 
remainder by phase~III, which can be obtained from the local maximum in the free 
energy (point \ahum{A} in \fig{fig1}(d)).

Since the interfaces are essentially flat, one can easily identify where the 
phases are located. In \fig{prof}(a), we see one large domain of phase I 
(characterized by a low overall magnetization) coexisting with one large domain 
of phase~III (characterized by large modulations in the magnetization). Hence, 
two I-III interfaces are present (recall that periodic boundaries are used). In 
\fig{prof}(b), we again observe I-III phase coexistence, but this time the 
phases are arranged such that four I-III interfaces are present. In equilibrium, 
arrangement (a) is preferred since it has the smallest interface area: $2L^2$ 
versus $4L^2$, with $L$ the lateral box size. However, for finite $L$, 
arrangement (b) is also frequently observed, since $\gamma$ is small. In fact, 
from the ratio of counts $R$, the interface tension can be determined
\begin{equation}\label{fit}
 \ln R = 2 \gamma L^2 + \Delta S, \quad R \equiv n_a/n_b,
\end{equation}
where $n_i$ denotes the number of times arrangement $i=(a,b)$ was seen during a 
very long simulation run (note that this simulation must be 
performed at fixed $m$ chosen to yield equal volumes of both phases). The 
\ahum{offset} $\Delta S$ reflects the combinatorial and translational entropy 
difference between the arrangements. The former is zero since there are as many 
ways to distribute the phases as in (a), as there are for (b). However, there is 
additional translational entropy for arrangement (a) since the domains are twice 
as large (we thus expect $\Delta S = \ln 2 \approx 0.69$).

To simulate at fixed $m$ we use Kawasaki dynamics: two spins of opposite sign 
are randomly selected and flipped, and the resulting spin configuration is 
accepted with the Metropolis criterion \cite{newman.barkema:1999}. To facilitate 
frequent transitions between arrangements (a) and (b) of \fig{prof}, we also use 
a collective Monte Carlo move. To this end, we introduce the block domain $B_i$, 
which contains all spins whose $z$-coordinate is between $\lambda b_i/2 < z \leq 
\lambda(b_i/2+1)$, with $b_i$ an integer (periodic boundary conditions must be 
applied). In the collective move, two block domains, $B_1$ and $B_2$, are 
randomly selected with the constraint that $|b_1 - b_2|>0$ and even. The domains 
are then swapped, and the resulting spin configuration is accepted with the 
Metropolis criterion (in our simulations, Kawasaki and collective moves are 
attempted in a ratio $1:0.03$, respectively).

To test our approach we consider $0.264 < J < 0.278$, which is between the 
triple and critical points. We use $m=-0.426$ for this is the value where phases 
I and III were seen to occupy equal volumes. In \fig{gam}(a), we plot $\ln R$ 
versus $L$ for $J=0.275$; the data are indeed well described by \eq{fit}, and by 
fitting $\gamma$ can be estimated. In \fig{gam}(b), we plot the corresponding 
estimates of $\gamma$ versus $J$. Despite the admittedly rather large 
statistical uncertainty, our data confirm that the tension is extremely low, and 
that it decreases as $J$ is lowered; both these observations are in qualitative 
agreement with theoretical predictions \cite{citeulike:9633508}.

\subsection{Correlations in the field direction}

\begin{figure}
\begin{center}
\includegraphics[width=0.9\columnwidth]{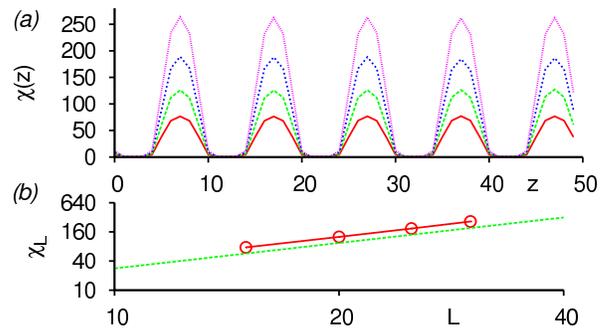}
\caption{\label{critcor} Investigation of the critical behavior using a 
simulation box with $D=5\lambda$; the simulations are performed at fixed 
magnetization $m=-0.426$ and $J=J_{\rm cr}$. (a) The susceptibility profiles 
$\chi(z)$ for $L=15,20,25,30$ (from bottom to top). The key point to note from 
this figure is that $\chi(z)$ diverges with $L$ only at special values~$z=z_{\rm 
cr}$. (b) Finite-size scaling analysis of the average peak height $\chi_L$ of 
the susceptibility profiles. We plot is $\chi_L$ versus $L$ on 
double-logarithmic scales. The dashed line corresponds to a power-law with 
exponent $\gamma/\nu=7/4$ of the $d=2$ Ising model.}
\end{center}
\end{figure}

\begin{figure}
\begin{center}
\includegraphics[width=0.9\columnwidth]{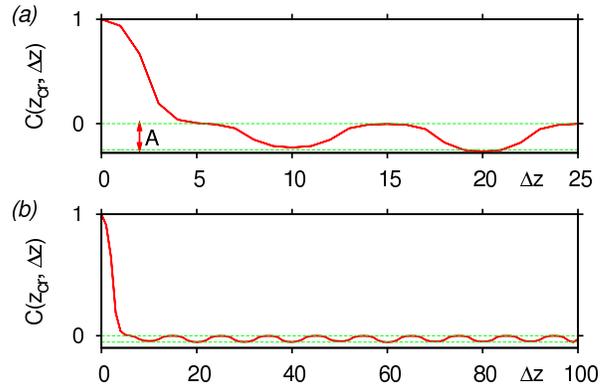}
\caption{\label{cfunc} The correlation function $C(z_{\rm cr},\Delta z)$ in the 
critical regime. We show results for $D=5\lambda, \, L=30$ (a), and 
$D=20\lambda, \, L=8$ (b). The vertical arrow in~(a) marks the amplitude $A$ of 
the anti-correlations, which conforms to \eq{eqamp}.}
\end{center}
\end{figure}

In the finite-size scaling analysis of \sect{sec:fss} we varied $L$ keeping 
$D=\lambda$ fixed. We thus assumed the correlations in the field direction to be 
short-ranged: critical correlations only develop in the lateral $L$~directions, 
but not in the direction $D$ along the field, such that the resulting critical 
behavior is effectively two-dimensional (and belonging to the $d=2$ Ising 
universality class). To verify this assumption we now consider the critical 
regime using a larger value $D=5\lambda$. We perform simulations at $J=J_{\rm 
cr}$ and fixed $m=-0.426$ (the latter corresponds to the average magnetization 
at the critical point). To simulate at fixed $m$ we use Kawasaki dynamics and 
collective moves (as in the previous section). However, for the collective 
moves, the block domain $B_i$ was taken to be a {\it single} lattice layer, 
containing those spins whose $z$-coordinate equals $z_i$ (at criticality, this 
choice yields a higher accept rate). A pair of layers is chosen randomly and 
swapped, with the constraint that the sign of $V_{\rm per}(z)$ in the layers is 
the same, and accepted with the Metropolis criterion.

In \fig{critcor}(a), we plot the susceptibility profile $\chi(z) = L^2 \left( 
\avg{m(z)^2} - \avg{m(z)}^2 \right)$ for various values of~$L$. The 
susceptibility diverges with $L$ only at selected values $z=z_{\rm cr}$, which 
\ahum{repeat} with the same period as the field. The critical behavior is thus 
spatially confined to those $L \times L$ slabs for which the corresponding 
$z$-coordinate equals one of the~$z_{\rm cr}$. To determine the universality 
class we compare the average peak heights $\chi_L$ of the susceptibility 
profiles to the finite-size scaling prediction $\chi_L \propto L^{\gamma/\nu}$, 
with $\gamma$ the susceptibility critical exponent. This result is shown in 
\fig{critcor}(b), and the $d=2$ Ising value $\gamma/\nu=7/4$ is strikingly 
confirmed. Hence, the observed universality class does not depend on the value 
of~$D$ used in the scaling analysis, which {\it a posteriori} provides the 
justification for the approach of \sect{sec:fss}.

Next, we ask whether correlations exist between critical slabs. To this end, we 
introduce the pair correlation function
\begin{multline}
 C(z_1,\Delta z) \propto \\
 \avg{m(z_1) \, m(z_1+\Delta z)} - 
 \avg{m(z_1)} \avg{m(z_1+\Delta z)},
\end{multline}
measured between the slab at $z=z_1$ and $z=z_1+\Delta z$, respectively. We 
choose $z_1$ to coincide with one of the critical slabs, and 
we normalize such that $C(z_1,0) \equiv 1$. In \fig{cfunc}, we show the 
correlation function for a system with $D=5\lambda$ (a), and for $D=20\lambda$ 
(b). We find that the slabs at $\Delta z = n \lambda$ with integer $n>0$ are 
{\it anti-correlated} from the (critical) slab at $n=0$. Moreover, the amplitude 
$A$ of the anti-correlations is independent of $\Delta z$, but it decreases with 
$D$. In fact, an almost perfect \ahum{lever rule} is observed
\begin{equation}\label{eqamp}
 A=\lambda/(D-\lambda). 
\end{equation}
That is: if there happens to be an excess magnetization in one of the critical 
slabs, the remaining critical slabs respond by assuming a lower magnetization, 
in a manner such that the excess magnetization is shared equally on average. In 
the limit $D \to \infty$, the amplitude $A$ of the correlations becomes zero, 
consistent with our assumption that long-ranged correlations in the field 
direction are absent. We also point out that the correlations in \fig{cfunc} are 
very different from critical correlations; the latter decay as power laws, 
$\lim_{\Delta z \to \infty} C(z_{\rm cr},\Delta z) \propto 1/\Delta z^\eta$ with 
critical exponent $\eta$, for which we see no evidence in our data. In fact, the 
anti-type correlations of \fig{cfunc} are also observed in the non-critical 
regime of the phase diagram (explicit checks were performed for $J=0.27$ 
using $m=-0.82$ and $m=-0.04$, corresponding to a pure phase I and 
phase III, respectively).

\section{Results in $d=2$ dimensions}

We now consider LIC in $d=2$ dimensions. The simulations are performed on $L 
\times D$ periodic lattices, with the field $V_{\rm per}(z)$ again propagating 
along edge $D$ of the lattice. In what follows, the field wavelength $\lambda=8$ 
with strength $h=0.1495$.

\subsection{Phase diagram and scaling analysis}

\begin{figure}
\begin{center}
\includegraphics[width=0.9\columnwidth]{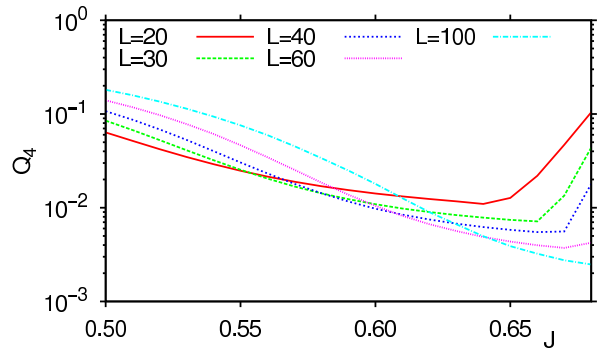}
\caption{\label{cr2d} The analogue of \fig{fss_cr}(a) but for the case $d=2$. 
Note the logarithmic vertical scale. The data are obtained using fixed 
$D=\lambda$. The key thing to note from this figure is that the curves for 
different $L$ do not intersect at a single point, implying the absence of a 
critical point. This, in turn, is consistent with $d=1$ Ising universality.}
\end{center}
\end{figure}

\begin{figure}
\begin{center}
\includegraphics[width=0.9\columnwidth]{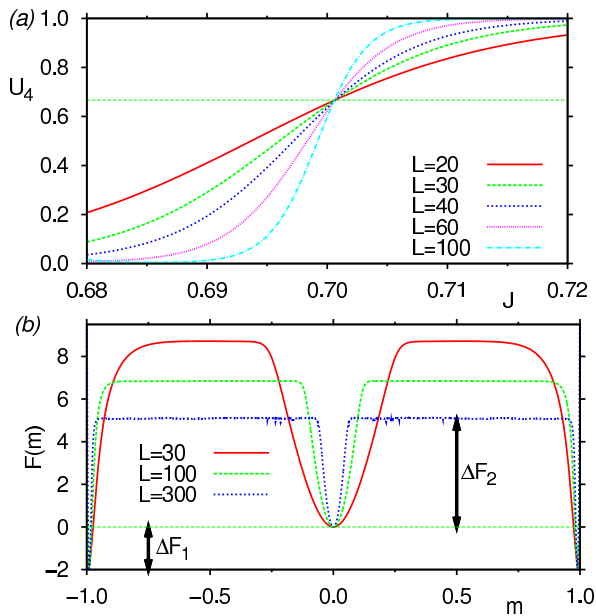}
\caption{\label{notrip} The analogue of \fig{fss_tr} but for the case $d=2$. The 
main difference is that we now observe a critical point, as opposed to a triple 
point. (a) The Binder cumulant as function of $J$ for $H=0$ and different system 
sizes $L$. The curves for different $L$ intersect from which we might conclude that a 
phase transition takes place. However, we display in (b) the scaling with $L$ of the free energy $F(m)$ 
at the cumulant intersection, with $F(m=0)$ shifted to zero, where we see that the depth of the 
central minimum $\Delta F_2 \to 0$ as $L$ increases, while the depth $\Delta 
F_1$ of the outer minima appears to be independent of $L$. This type of scaling 
is consistent with a critical point \cite{citeulike:3908342}.}
\end{center}
\end{figure}

We first determine whether the LIC critical points occur in $d=2$ dimensions 
also. Since the critical behavior was shown to resemble that of a reduced 
dimension $d-1$, we now expect the universality class of the $d=1$ Ising model. 
As is well known, the latter model does not feature a critical point. In 
\fig{cr2d}, we repeat the cumulant analysis of \fig{fss_cr}(a). In line with the 
$d=1$ Ising model, we do not observe an intersection point, confirming the 
absence of a critical point. While for small~$L$ the curves somewhat intersect, 
the intersections for larger $L$ systematically shift toward larger values 
of~$J$. Hence, in $d=2$ dimensions, there is no LIC critical behavior.

Next, we investigate the fate of the triple point, using the same analysis as in 
\fig{fss_tr}. We collect data for fixed $H=0$ and $D=\lambda$, while $J$ and $L$ 
are varied. In \fig{notrip}(a), we plot the Binder cumulant $U_4$ versus $J$ for 
different system sizes $L$. Consistent with a triple point, we observe a sharp 
intersection, with the value of the cumulant at the intersection very close to 
$U_4=2/3$ of a triple-peaked distribution. However, the corresponding free 
energy is {\it not} consistent with a triple point, see \fig{notrip}(b), where 
$F(m)$ is plotted for three different system sizes; note that we plot $F(m)$ 
with the central ($m=0$) minimum shifted to zero. While $F(m)$ clearly reveals 
three minima, the central minimum does not survive in the thermodynamic limit. 
This can be seen from the corresponding \ahum{depth}, marked $\Delta F_2$ in the 
figure, which decreases with~$L$. In the limit $L \to \infty$, we have $\Delta 
F_2 \to 0$, and only the outer minima survive, whose corresponding depths then 
equal $\Delta F_1$. The observation in \fig{notrip}(b) that $\Delta F_1$ is 
independent of system size is characteristic of a continuous transition 
\cite{citeulike:3908342}. Hence, for LIC in $d=2$ dimensions, the triple point 
is destroyed, and replaced by a critical point, in this case at $J_{\rm cr} 
\approx 0.701$ (as expected, this exceeds $J_{\rm cr,bulk} = \ln(1+\sqrt{2})/2$ 
of the bulk $d=2$ Ising model). The LIC phase diagram in $d=2$ dimensions is 
thus radically different from $d=3$. Instead of a \ahum{pitchfork} topology, we 
now have a single line of first-order phase transitions terminating in a 
critical point (\fig{fig2}(b)).

\begin{figure}
\begin{center}
\includegraphics[width=0.9\columnwidth]{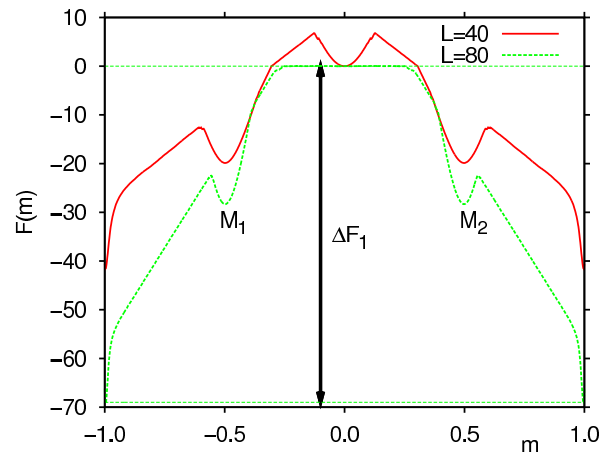}
\caption{\label{2d1} Free energy $F(m)$ for $d=2$ and $J = 0.8 > J_{\rm cr}$; 
the data are obtained using $D=2\lambda$, $H=0$, and two values of $L$ as 
indicated. The key thing to note from this figure is that the free 
energy barrier $\Delta F_1$ increases with $L$, indicating a first-order phase 
transition \cite{citeulike:3908342}. Note also the spurious minima $M_1$ and 
$M_2$: these reflect meta-stable coexistence states \cite{citeulike:9633508} 
whose role in the thermodynamic limit is negligible.}
\end{center}
\end{figure}

\begin{figure*}
\begin{center}
\includegraphics[width=2\columnwidth]{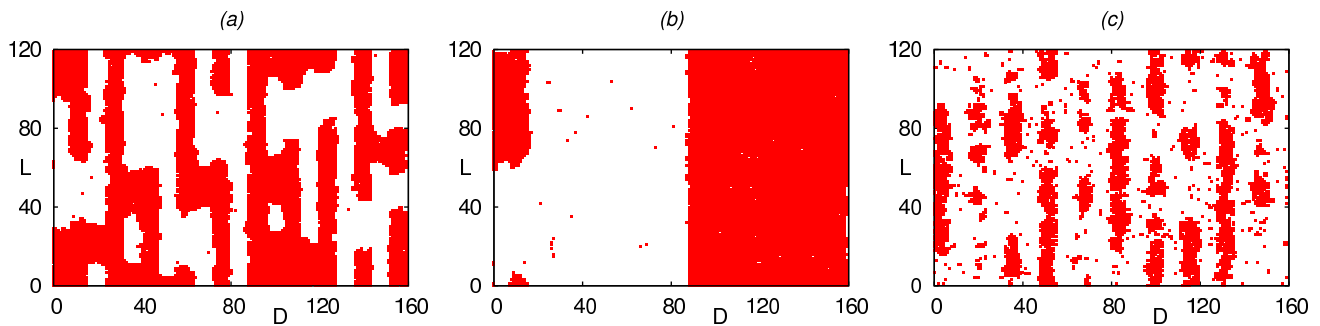}
\caption{\label{snap} Computer generated snapshots obtained using $D=10\lambda$ 
and $L=120$; white regions correspond to \ahum{spin-up}. The snapshots in (a) 
and (b) are obtained for $J=0.8$ and $m=0$ and show \ahum{early time} and \ahum{late 
time} snapshots, respectively (the simulation was started with a random spin 
configuration). Since $J>J_{\rm cr}$ we observe coarsening of domains (a) until 
two large domains have formed (b). In (c), we show a snapshot for the case when $J=0.5<J_{\rm 
cr}$ and $m=0.52$. In this situation, domains do not coarsen with time, but remain 
finite in size, reminiscent of the $d=1$ Ising model.}
\end{center}
\end{figure*}

We still find that, for $J>J_{\rm cr}$, the transition is first-order. In 
\fig{2d1}, we plot the free energy $F(m)$ for $J=0.8, \, H=0$ using system sizes 
$L=40,80$ and $D=2\lambda$. The free energy curves are again shifted such that 
$F(m=0)=0$. While for the smaller system the minimum at $m=0$ is still visible, 
it has vanished in the larger system. In addition, the barrier $\Delta F_1$ now 
increases profoundly with~$L$, consistent with a first-order transition 
\cite{citeulike:3908342}. Note also the pronounced flat region in $F(m)$ around 
$m \sim 0$ for the larger system: this indicates two-phase coexistence with 
negligible interactions between the interfaces \cite{citeulike:7237424}. When a 
simulation is performed in this regime starting from a random initial spin 
configuration, the system phase separates to form structures that are strongly 
affected by the external potential; see Fig.\ \ref{snap}(a). However, when the 
system is fully equilibrated at a \ahum{later time}, snapshots show the system 
containing two coexisting domains of phases I and II; see Fig.\ \ref{snap}(b). 
In a box with periodic boundaries, the domains arrange themselves as two slabs 
since this minimizes the total interface length.

\subsection{Rounding effects}

Even though the \ahum{zebra} phase (phase III), does not survive in the 
thermodynamic limit in $d=2$, we still see remnants of this phase in systems 
of finite size. If one simulates at $J<J_{\rm cr}$ using an appropriate external 
field~$H$, one finds that in finite systems $F(m)$ can still be cast into the 
forms of \fig{fig1}(d,e). Inspection of simulation snapshots then reveals a 
condensation of droplets onto stripes oriented perpendicular to the field 
direction (\fig{snap}(c)). However, the droplet size remains finite in this 
case, owing to the fact that the $d=1$ Ising model at finite temperature does 
not support a finite magnetization. Similar finite-size effects occur in 
colloid-polymer mixtures confined to cylindrical pores, which also belong to the 
universality class of the $d=1$ Ising model \cite{citeulike:7678249, 
citeulike:8132587}.

\section{Summary}

In this work, we considered the phase behavior of the Ising model exposed to a 
static periodic field. In $d=3$ dimensions, we obtain a phase diagram analogous 
to laser-induced condensation observed in fluids undergoing bulk \LV type 
transitions. That is, a new phase arises (the \ahum{zebra} phase) and the 
critical point of the bulk model is replaced by two new critical points, and a 
triple point. The main difference compared with fluids is that, due to spin reversal 
symmetry, the corresponding phase diagram for the Ising model features a symmetry 
line. The analysis of the present work complements earlier works on 
laser-induced condensation \cite{citeulike:7530595, citeulike:9633508} in that 
(i) a detailed study of finite-size effects at the triple point was presented, 
(ii) a simple mean-field theory was used to elucidate in a qualitative manner
how the $d=3$ phase transitions depend on the parameters in the external potential, (iii)
a method was presented to measure the extremely low tension of interfaces 
with the zebra phase, and (iv) the nature of correlations along the field 
direction was further clarified.

We additionally considered the fate of laser-induced condensation in $d=2$ 
dimensions. In this case, we find that the zebra phase does not survive in the 
thermodynamic limit, and the corresponding phase diagram features just a single 
critical point. This critical point occurs at a temperature below the critical 
temperature of the pure $d=2$ Ising model. The universality class of the 
critical point still needs to be determined. The analysis of the free energy in 
\fig{notrip}(b) only indicates a critical transition, since the barrier $\Delta 
F_1$ is $L$-independent, but no information regarding critical exponents could 
be obtained. The practical problem here is that, in computer simulations, we are 
still restricted to system sizes that span only a few field wavelengths.
We should also mention that the mean-field theory used in Sec.\ III B predicts
very similar results in $d=2$ as it does in $d=3$ and is therefore not reliable when applied
in $d=2$.

Even though our results were obtained for the relatively simple Ising model, the 
generic features of the observed phase behavior should also apply to real fluids. 
In particular the experimental realization in $d=2$ dimensions should be 
feasible using a stripe-patterned substrate. At moderate temperatures, the 
condensation of finite-sized droplets should be observable, while at low 
temperatures a macroscopic demixing should occur (c.f.\ \fig{snap}).

\acknowledgments

RLCV was supported by the {\it Deutsche Forschungsgemeinschaft}
(Emmy Noether program:~VI~483/1-1) and AJA was supported by RCUK.

\newpage

\bibliography{refs}

\end{document}